\documentclass[aps,prb,twocolumn,floatfix,superscriptaddress]{revtex4-2}
\usepackage[utf8]{inputenc}
\usepackage[T1]{fontenc}
\usepackage[english]{babel}
\usepackage{mathtools}
\usepackage{amsmath}
\usepackage{amsfonts}

\usepackage{booktabs}

\usepackage[dvipsnames,svgnames,x11names,hyperref]{xcolor}
\usepackage{hyperref}
\hypersetup{%
    colorlinks=true, linkcolor=NavyBlue, urlcolor=NavyBlue, citecolor=NavyBlue
}

\usepackage{physics,braket,siunitx,slashed}
\usepackage[capitalise]{cleveref}

\usepackage{tabularx}

\newcommand{\bk}{\mathbf{k}}
\newcommand{\br}{\mathbf{r}}
\newcommand{\bG}{\mathbf{G}}
\newcommand{\bQ}{\mathbf{Q}}
\newcommand{\bq}{\mathbf{q}}

\begin{document}

\title{Roton softening and crystallization in a mass-asymmetric electron--hole bilayer}

\author{Ambuj Jain}
\affiliation{Department of Physics and Astronomy, University of Kentucky, Lexington, Kentucky 40506-0055, USA}

\author{Mu-Yang Chen}
\affiliation{Department of Physics, The Pennsylvania State University, University Park, Pennsylvania 16802, USA}

\author{Chao-Xing Liu}
\affiliation{Department of Physics, The Pennsylvania State University, University Park, Pennsylvania 16802, USA}
\affiliation{Center for Theory of Emergent Quantum Matter,
The Pennsylvania State University, University Park, Pennsylvania 16802, USA}

\author{Binghai Yan}
\affiliation{Department of Physics, The Pennsylvania State University, University Park, Pennsylvania 16802, USA}
\affiliation{Center for Theory of Emergent Quantum Matter,
The Pennsylvania State University, University Park, Pennsylvania 16802, USA}

\author{Wangqian Miao}
\affiliation{Department of Physics, The Pennsylvania State University, University Park, Pennsylvania 16802, USA}
\affiliation{Center for Theory of Emergent Quantum Matter,
The Pennsylvania State University, University Park, Pennsylvania 16802, USA}

\date{\today}

\begin{abstract}
Whether roton softening drives a continuous transition into a density-ordered or supersolid state is a longstanding question in interacting quantum fluids. Here, we address this
microscopically in a mass-asymmetric electron--hole bilayer in which exciton condensation competes with interaction-driven crystallization. Using self-consistent Hartree--Fock--Bogoliubov theory and its time-dependent extension, we show that  increasing the carrier-mass ratio or interlayer distance induces a pronounced roton minimum in the homogeneous exciton condensate liquid. The system undergoes a first-order transition to a crystalline state, preempting the roton instability. In this metastable regime, strongly interacting holes first drive the crystallization of the itinerant electrons, and the system then evolves into a hole-crystal--electron-metal phase at larger interlayer distance. Supersolid solutions with coexisting interlayer coherence and charge density order are also found to be energetically competitive near the crystallization boundary. We further discuss our findings based on recent experimental setups in two-dimensional material heterostructures.
\end{abstract}
\maketitle

\section{Introduction}

Spatially separated electron--hole bilayers have emerged as a robust platform for realizing equilibrium excitonic matter~\cite{DePaloRapisardaSenatore2002,MaezonoEtAl2013,LopezRiosEtAl2018,ShaoDai2024QO,ShaoDai2024Breakdown}. Recent observations of strongly correlated electron--hole fluids and nearly perfect Coulomb drag~\cite{WangEtAl2019Nature,MaEtAl2021Nature,QiEtAl2023Thermodynamic,NguyenEtAl2025Science,QiEtAl2025Science} establish the formation of strongly locked dipolar excitons, shifting the central question from their experimental realization to the nature of these exotic quantum phases at low temperatures.

An important next frontier is interaction-driven crystallization. A stable exciton crystal was recently observed in a hybrid electron--hole system formed by a $\mathrm{MoSe}_{2}$ monolayer coupled to a $\mathrm{WS}_{2}/\mathrm{WSe}_{2}$ moir\'e superlattice~\cite{QiEtAl2026ExcitonCrystal}. The kinetic energy of the holes on the moir\'e side is strongly quenched, which favors localization and traps the electrons on the other side. This crystalline state was identified through optical Umklapp scattering and exciton transport at a commensurate filling. However, the underlying moir\'e potential explicitly selects the ordering wave vector and pins the crystalline orientation. It therefore remains unclear whether mass asymmetry and Coulomb interactions alone are sufficient to drive spontaneous crystallization in a translationally invariant electron--hole bilayer.

\begin{figure}[t]
    \centering
    \includegraphics[width=\columnwidth]{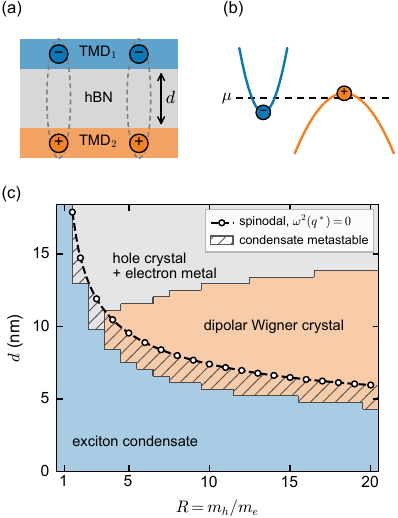}
    \caption{
        Schematic of a mass-asymmetric electron--hole bilayer.
        (a) Two transition-metal dichalcogenide (TMD) materials, TMD$_1$ and TMD$_2$, are separated
        by a hexagonal boron nitride (hBN) spacer of thickness $d$. The dashed ellipses denote
        interlayer excitons.
        (b) Corresponding band alignment at charge balance. The more
        dispersive electron band and flatter hole band represent
        $m_e<m_h$. The dashed line denotes the common chemical
        potential $\mu$ alignment, which can be tuned by an external gate.
        (c) Phase diagram at fixed electron coupling $r_s^e=2$ ($n=10^{11}$~cm$^{-2}$, $m_e=0.07\,m_0$, $\varepsilon_r=11.8$) versus mass ratio $R=m_h/m_e$ and layer separation $d$. Colored cells mark the HFB ground state: exciton condensate (blue), dipolar Wigner crystal (orange), and hole crystal coexisting with an electron metal (gray); the dipolar crystal first appears at $R=4$ on our integer-$R$ grid. The dashed line is the spinodal of the uniform condensate, $\omega^2(q^*)=0$ with $q^*\simeq2k_F\approx G_1$. In the hatched region the condensate survives as a metastable state.}
    \label{fig:mass_asymmetric_bilayer}
\end{figure}

The central theoretical question is how a homogeneous exciton condensate loses stability. Increasing the interlayer separation or the mass ratio enhances the dipolar repulsion relative to the kinetic energy. For dilute dipolar excitons, this produces a roton minimum in the collective spectrum~\cite{FedorovKurbakovLozovik2014,FilShevchenko2016,FilShevchenko2021}. Within time-dependent Hartree--Fock (TDHF) theory, the uniform condensate supports Goldstone and Higgs-like modes~\cite{XueWuMacDonald2020,ShaoShiDai2025}.
In a perpendicular magnetic field, the Goldstone mode develops a magnetoroton minimum that signals stripe order~\cite{ShaoShiDai2025}.
Roton softening has consequently been proposed as a precursor to finite-momentum condensation or an excitonic supersolid~\cite{JoglekarBalatskyDasSarma2006,ContiEtAl2023Chester,ContiEtAl2026GPSupersolid,DaiFu2024Trions}. The key question, however, is not simply whether a roton can develop, but what phase actually replaces the uniform condensate as the roton deepens.

Whether roton softening produces a supersolid is contested. A coherent-state mean-field theory of the dilute limit finds a first-order transition to a hexagonal supersolid just before the roton softens~\cite{FilShevchenko2020}. Path-integral Monte Carlo instead finds that the exciton superfluid crystallizes with an abrupt loss of coherence~\cite{BoeningFilinovBonitz2011}. Separately, path-integral Monte Carlo~\cite{BonitzEtAl2005,LudwigEtAl2007,SchleedeEtAl2012}, STLS~\cite{AroraMoudgil2021}, and Hartree--Fock~\cite{NashabehFu2026} studies of mass-asymmetric systems established three regimes: an exciton condensate at small separation, a heavy-hole crystal coexisting with a light-electron liquid at large mass ratio, and crystalline phases in between~\cite{ChaiEtAl2026Honeycomb}. Quantum Hall bilayers provide a related setting in which kinetic energy is quenched: Hartree--Fock theory finds roton softening together with phases exhibiting simultaneous interlayer coherence and density modulation~\cite{Fertig1989,BreyFertig2000,CoteFertig2002,DasSarmaChou2026}. Whether such coexistence survives at zero field, where the kinetic energy is not quenched, remains open. What is missing is a single microscopic framework that treats interlayer coherence, crystallization, and the collective modes of the resulting phases on equal footing. Such a framework can determine what the rotonized condensate freezes into, how the roton-instability line is related to the thermodynamic phase boundary, and whether the competing crystalline states are dynamically stable.

In this work, we study a density-balanced, mass-asymmetric electron--hole bilayer at zero temperature and zero magnetic field using self-consistent Hartree--Fock--Bogoliubov (HFB) theory and its time-dependent extension. The bilayer is characterized by the mass ratio $R=m_h/m_e$, the interlayer separation $d$, and the Wigner--Seitz radii $r_s^{e,h}=1/(\sqrt{\pi n}\,a_B^{e,h})$ at the common density $n$. Here $a_B^{e,h}$ are the effective Bohr radii, so $r_s^h=Rr_s^e$. We fix $r_s^e=2$ ($n=10^{11}$~cm$^{-2}$), for which an isolated electron layer remains well on the liquid side of the Wigner-crystallization regime. In the Nambu basis of electrons and holes, a single folded plane-wave HFB formulation describes the uniform condensate, density order in either layer, and their coexistence, and its linearization gives the collective modes of every phase.

Starting from the uniform exciton condensate, we find that increasing either the mass ratio $R$ or the interlayer separation $d$ produces a pronounced roton minimum near $2k_F$, close to the ordering wave vector of the competing triangular crystal. Crystallization, however, occurs before the roton mode approaches zero: the ground state undergoes a first-order transition from the uniform condensate to a crystalline state without interlayer coherence, while the condensate remains locally stable up to a distinct roton spinodal. Thus the thermodynamic transition and the dynamical instability are separated by a metastable condensate regime. Within the commensurate sector with one electron and one hole per unit cell, the crystalline state remains energetically favored over states with simultaneous coherence and density order. Upon relaxing this filling constraint, we find supersolid solutions with coexisting interlayer coherence and crystalline order that become energetically competitive near the phase boundary.

At sufficiently large mass asymmetry, the crystalline phase first appears as a dipolar Wigner crystal with density order in both layers. The heavy-hole lattice induces density modulation in the itinerant electron layer, while increasing the interlayer separation weakens this locking and eventually produces a hole crystal coexisting with an electron metal. The ordered state also supports acoustic and optical collective modes, with the optical branches strongly softened relative to their classical bilayer counterparts. Together, these results connect roton softening, metastability, supersolid competition, and spontaneous crystallization within a single microscopic framework. They also provide a reference point for interaction-driven excitonic crystals beyond moir\'e-assisted settings.

\section{Model and Method}\label{sec:model}

\subsection{Model setup}

We first introduce the continuum Hamiltonian for the electron--hole bilayer. Working in the electron and hole representation, we use $e_{\bk}$ and $h_{\bk}$ to denote electrons in the conduction band of the electron layer and holes in the valence band of the hole layer, respectively:
\begin{equation}
\label{eq:Hamiltonian}
\begin{aligned}
H ={}& \sum_{\bk} \left[
\left( \frac{\hbar^2 k^2}{2m_e} - \mu_e \right) e_{\bk}^{\dagger}e_{\bk}
+ \left( \frac{\hbar^2 k^2}{2m_h} - \mu_h \right) h_{\bk}^{\dagger}h_{\bk} \right] \\
&+ \frac{1}{2S} \sum_{\bk,\bk',\bq} v_0(q) e_{\bk+\bq}^{\dagger} e_{\bk'-\bq}^{\dagger} e_{\bk'} e_{\bk} \\
&+ \frac{1}{2S} \sum_{\bk,\bk',\bq} v_0(q) h_{\bk+\bq}^{\dagger} h_{\bk'-\bq}^{\dagger} h_{\bk'} h_{\bk} \\
&- \frac{1}{S} \sum_{\bk,\bk',\bq} v_1(q) e_{\bk+\bq}^{\dagger} h_{\bk'-\bq}^{\dagger} h_{\bk'}e_{\bk}.
\end{aligned}
\end{equation}
Here,
\begin{equation}
v_0(q) = \frac{e^2}{2\varepsilon_0\varepsilon_r q}, \qquad v_1(q) = v_0(q)e^{-qd},
\end{equation}
are the intra- and interlayer Coulomb interactions, respectively, $d$ is the interlayer separation, and $S$ is the system area. For simplicity, we suppress spin and valley indices, assuming a fully spin- and valley-polarized state. The hole mass $m_h>0$ is set by the curvature at the valence-band maximum, $\varepsilon_v(\bk)\simeq-\hbar^2k^2/2m_h$, and can serve as an effective parameter incorporating moir\'e flattening effects, as discussed below. In this representation the intralayer electron--electron and hole--hole interactions are repulsive, while the interlayer electron--hole interaction is explicitly attractive, as reflected in the sign of the last term of Eq.~\eqref{eq:Hamiltonian}. The Hamiltonian describes the heterostructure shown in Fig.~\ref{fig:mass_asymmetric_bilayer}(a), in which the electron and hole layers are separated by a hexagonal boron nitride (hBN) spacer that suppresses direct tunneling and interlayer recombination. The chemical potentials $\mu_e$ and $\mu_h$ are chosen to yield equal electron and hole densities, $n_e=n_h=n$, so that the Fermi wave vectors match, $k_F^e=k_F^h=\sqrt{4\pi n}$, as illustrated in Fig.~\ref{fig:mass_asymmetric_bilayer}(b). We define the mass ratio $R\equiv m_h/m_e$.

The interaction strength for each layer $\alpha=e, h$ is parameterized by the Wigner--Seitz radius $r_s^\alpha = 1/(\sqrt{\pi n} a_B^\alpha)$, where the effective Bohr radius is given by $a_B^\alpha = 4\pi\varepsilon_0\varepsilon_r\hbar^2/m_\alpha e^2$. At equal densities, $r_s^h=R\,r_s^e$. Recent experiments realize several complementary regimes for these parameters. Continuum MoSe$_2$/hBN/WSe$_2$ bilayers operate in the regime $r_s^e \simeq 21\text{--}60$ and $r_s^h \simeq 12\text{--}34$ over the experimentally relevant density range, exhibiting only a moderate bare mass asymmetry~\cite{NguyenEtAl2025Science,QiEtAl2025Science}. In the MoSe$_2$/hBN/WS$_2$/WSe$_2$ exciton-crystal device, the narrow moir\'e valence band strongly suppresses the hole kinetic energy, making its effective $r_s^h$ substantially larger than the estimate based on the bare WSe$_2$ mass~\cite{QiEtAl2026ExcitonCrystal}.

\subsection{Hartree--Fock--Bogoliubov theory }
The model Hamiltonian, Eq.~\eqref{eq:Hamiltonian}, admits two distinct spontaneous symmetry breakings. The attractive interlayer interaction $v_1$ favors electron--hole pairing, producing an exciton condensate that breaks the $U(1)$ symmetry generated by $N_e+N_h$. The intralayer repulsion $v_0$ favors crystallization, producing a Wigner crystal that breaks continuous translational symmetry down to a discrete lattice symmetry. These orders are not independent. The interlayer interaction that drives pairing also couples the charge densities of the two layers, so that a crystal in the heavier layer can induce crystalline order in the lighter one. We therefore treat both on equal footing within a self-consistent Hartree--Fock--Bogoliubov (HFB) calculation, allowing the mean-field ansatz to develop either or both symmetry breakings spontaneously.

To accommodate both orders, we take the variational state to be the most general HFB quasiparticle vacuum invariant under translations of a triangular lattice. We assume that the self-consistent crystalline order has the same periodicity in both layers; since the densities are equal, both layers then have the same filling fraction $\nu$, the number of particles per unit cell. The lattice constant $a=[2\nu/(\sqrt{3}n)]^{1/2}$ is then fixed by $n$ and $\nu$. The number of $\bk$ points $N_k$ equals the number of unit cells, so $N_e=N_h=\nu N_k$. Momenta are decomposed as $\bk+\bG$, with $\bk$ in the Brillouin zone and $\bG$ a reciprocal-lattice vector. We introduce the Nambu spinor
\begin{equation}
  \Psi_{\bk}=\big(e_{\bk+\bG_1},\dots,e_{\bk+\bG_{N_G}},\,
  h^{\dagger}_{-\bk+\bG_1},\dots,h^{\dagger}_{-\bk+\bG_{N_G}}\big)^{T},
  \label{eq:nambu}
\end{equation}
where $N_G$ is the number of retained reciprocal-lattice vectors. The state is fully specified by the $2N_G\times 2N_G$ Hermitian generalized density matrix $\Gamma_{ab}(\bk)=\langle\Psi^{\dagger}_{\bk,b}\Psi_{\bk,a}\rangle$,
\begin{equation}
  \Gamma(\bk)=
  \begin{pmatrix}
    \rho^{e}(\bk) & \Phi(\bk)\\
    \Phi^{\dagger}(\bk) & 1-\bar{\rho}^{h}(\bk)
  \end{pmatrix},
  \label{eq:rho}
\end{equation}
with $\rho^{\alpha}_{\bG\bG'}(\bk)=\langle\alpha^{\dagger}_{\bk+\bG'}\alpha_{\bk+\bG}\rangle$ for $\alpha\in\{e,h\}$, $\bar\rho^{h}_{\bG\bG'}(\bk)=\langle h^{\dagger}_{-\bk+\bG}h_{-\bk+\bG'}\rangle=\rho^{h}_{\bG'\bG}(-\bk)$, and $\Phi_{\bG\bG'}(\bk)=\langle h_{-\bk+\bG'}e_{\bk+\bG}\rangle$.

The order parameters are read off directly from $\Gamma$. The normal blocks give the density Fourier components $\tilde{n}_\alpha(\bG)=N_k^{-1}\sum_{\bk\bG'}\rho^{\alpha}_{\bG'+\bG,\bG'}(\bk)$, normalized so that $\tilde{n}_\alpha(0)=\nu$, which are nonzero at $\bG\neq0$ only in a crystal. As the crystalline order parameter of layer $\alpha$ we use the first-shell amplitude $\langle|\tilde{n}_\alpha(\bG_1)|\rangle$, the average of $|\tilde{n}_\alpha(\bG)|$ over the six shortest reciprocal-lattice vectors. Since $n_\alpha(\br)\ge0$, $|\tilde{n}_\alpha(\bG)|\le\tilde{n}_\alpha(0)$, with equality only for point particles on the lattice sites; the order parameter therefore runs from $0$ in a uniform layer to $1$ in a perfectly localized crystal. The anomalous block $\Phi$ is the condensate amplitude, reducing to $\Phi_{\bk}=\langle h_{-\bk}e_{\bk}\rangle$ in a translation-invariant condensate. The ansatz imposes no relation between $n_e(\bG)$ and $n_h(\bG)$, so the relative displacement of the two crystals, like the phase structure of $\Phi$, is an output of the self-consistency. The only restriction on pairing is that the pair momentum vanishes modulo reciprocal-lattice vectors. $\Phi$ may therefore be modulated at the lattice wave vectors, with components $\langle h_{-\bk}e_{\bk+\bG}\rangle$, whereas pairing at an incommensurate momentum $\bQ\notin\{\bG\}$ (FFLO type) is excluded. Since $k_F^e=k_F^h$ for any $R$, the electron and hole Fermi contours are nested at zero pair momentum, and mean-field pairing remains at $\bQ=0$ also for unequal masses~\cite{ZhuDasSarma2025}; finite-momentum pairing requires a large density imbalance~\cite{ParishMarchettiLittlewood2011,PieriNeilsonStrinati2007}.

The interactions enter the self-consistent field through three channels. Both $v_0$ and $v_1$ contribute to the Hartree field $\Sigma^{H}$, which acts within the normal blocks and depends on $\Gamma$ only through $n_\alpha(\bG)$, with $\bG\neq0$. The intralayer interaction $v_0$ alone generates the Fock field $\Sigma^{F}$, which also acts within the normal blocks since $e$ and $h$ are distinct species. The interlayer interaction $v_1$ alone generates the pairing field $\Sigma^{X}$, linear in $\Phi$, which occupies the anomalous blocks. The separation $d$ therefore enters the two orders differently. It affects the condensate directly through $\Sigma^{X}$, while it affects the crystalline order only through the interlayer Hartree term. The energy of the converged state is $E=E^{\rm kin}+E^{H}+E^{F}+E^{X}$, with
\begin{subequations}
\label{eq:energy}
\begin{align}
  E^{H}+E^{F}&=\frac{1}{2}\sum_{\bk,\alpha}\mathrm{Tr}\big[(\Sigma^{H}+\Sigma^{F})_{\alpha}(\bk)\,\rho^{\alpha}(\bk)\big],
  \label{eq:energy_HF}\\
  E^{X}&=\sum_{\bk}\mathrm{Tr}\big[\Sigma^{X}(\bk)\Phi^{\dagger}(\bk)\big].
  \label{eq:energy_X}
\end{align}
\end{subequations}
We quantify interlayer coherence by the fraction $E^{X}/E_{\rm int}$, where $E_{\rm int}=E^{H}+E^{F}+E^{X}$.

At each $(R,d)$ we solve the self-consistent equations with several seed states, namely a uniform condensate, a uniform liquid, a crystal in both layers with $\Phi=0$, a crystal with interlayer coherence, and a random $\Gamma$, and take the lowest-energy converged solution as the ground state. Numerical details are given in Appendix~\ref{app:hf}.

\section{Results}\label{sec:results}

\subsection{Phase diagram at commensurate filling}
Figure~\ref{fig:mass_asymmetric_bilayer}(c) is the zero-temperature phase diagram of the density-balanced electron--hole bilayer as a function of the mass ratio $R=m_h/m_e$ and interlayer separation $d$, at fixed $r^e_s = 2$. Increasing $R$ strengthens the hole-layer coupling $r_s^h=Rr_s^e$, while increasing $d$ weakens the interlayer interaction. We find three phases: a uniform exciton condensate (EC), a dipolar Wigner crystal (dWC) with density order in both layers, and a hole crystal coexisting with an electron metal (HC+EM), classified by the criteria of Table~\ref{tab:criteria}. The diagram is computed at one electron and one hole per unit cell, $\nu=1$.

At small $d$ the ground state is a uniform exciton condensate: interlayer coherence is finite and both layers are uniform, $\langle|\tilde{n}_{e,h}(\bG_1)|\rangle<10^{-2}$. The condensate breaks the $U(1)$ symmetry generated by $N_e+N_h$, which fixes the relative phase of the two layers. For $R=1$ the EC is the ground state throughout the range of $d$ we considered. With increasing $R$ its region shrinks, the boundary falling from $d\approx13.0$~nm at $R=2$ to $4.3$~nm at $R=20$, as heavier holes favor crystalline order. The dashed line marks where the uniform condensate loses local stability. In the hatched region between the phase boundary and this line, the condensate survives as a metastable state above the crystalline ground state. The relation between this stability limit and the equilibrium transition is examined in the following subsection.

In both the dWC and the HC+EM phase the hole layer is an insulating crystal; the two phases differ only in the state of the electron layer, which is always modulated at the hole-lattice wave vectors. For $R\geq4$, an intermediate window of $d$ hosts the dWC, in which the electron layer is also gapped. This window widens with increasing $R$. At larger $d$ the hole-crystal potential acting on the electrons weakens, the electron bands overlap, and the electron layer becomes a compensated semimetal with only a weak induced density modulation, which defines the HC+EM phase. For $R=2$ and $3$ the EC gives way directly to HC+EM. The dWC--HC+EM boundary is therefore a metal--insulator transition of the electron layer within a hole crystal.

\begin{table}
\begin{ruledtabular}
\begin{tabular}{lccccc}
Phase & $E^X/E_{\rm int}$ (\%) & $\bar{n}_{G_1}$ & $\Delta_e$ & $\Delta_h$ & $\nu$ \\
\hline
EC               & $> 10^{-1}$ & $< 10^{-2}$ & ---        & ---        & ---      \\
Supersolid       & $> 10^{-1}$ & $> 10^{-2}$ & ---        & ---        & any      \\
dWC              & $< 10^{-1}$ & $> 10^{-2}$ & $> \delta$ & $> \delta$ & $1$      \\
HC+EM            & $< 10^{-1}$ & $> 10^{-2}$ & $< \delta$ & $> \delta$ & $1$      \\
Metallic crystal & $< 10^{-1}$ & $> 10^{-2}$ & metallic   & metallic   & $\neq 1$ \\
Liquid           & $< 10^{-1}$ & $< 10^{-2}$ & metallic   & metallic   & ---      \\
\end{tabular}
\end{ruledtabular}
\caption{Classification of mean-field states. $E^X/E_{\rm int}$ is the percentage of the interaction energy carried by interlayer coherence, $\bar{n}_{G_1}=\max_\alpha\langle|\tilde{n}_\alpha(\bG_1)|\rangle$ is the larger first-shell density Fourier amplitude of the two layers ($0\le\bar{n}_{G_1}\le1$), and $\Delta_\alpha$ is the indirect quasiparticle gap of layer $\alpha$, with $\delta=0.1$~meV. A dash means the quantity does not enter the classification. At $\nu\neq1$ the filling per unit cell is non-integer, such crystalline state has a fermi surface, so its a metallic crystal.}
\label{tab:criteria}
\end{table}

\subsection{Roton instability}

\begin{figure}[t]
  \includegraphics[width=\linewidth]{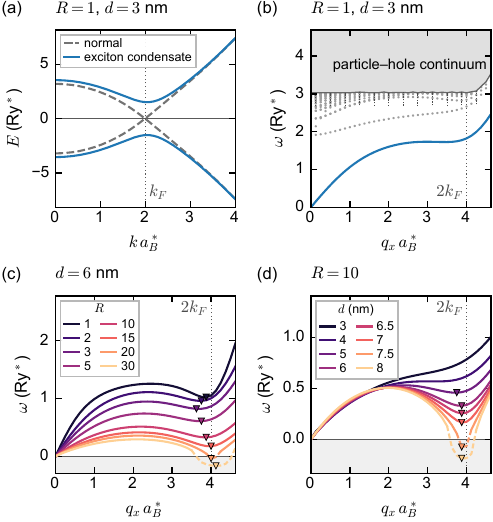}
    \caption{Roton softening in a bilayer exciton condensate at $n=10^{11}$~cm$^{-2}$ and $r_s^e=2$ ($m_e=0.07\,m_0$, $\varepsilon_r=11.8$), with mass ratio $R=m_h/m_e$. All panels use fixed excitonic units defined by the $R=1$ reduced mass $\mu_{\mathrm{ref}}=m_e/2$: $a_B^*=2a_B^e=17.84$~nm and $\mathrm{Ry}^*=3.42$~meV. (a)~HFB quasiparticle bands at $R=1$, $d=3$~nm, with (solid) and without (dashed) the off-diagonal condensate field, keeping the diagonal fields fixed. (b)~TDHF spectrum along $\bq=(q_x,0)$ at the same parameters. Dots show the computed levels, shading the particle--hole continuum, and the blue line the Goldstone branch. (c)~Goldstone branch for varying $R$ at $d=6$~nm. (d)~Goldstone branch for varying $d$ at $R=10$. Dotted lines mark $k_Fa_B^*=2$ in (a) and $2k_Fa_B^*=4$ in (b)--(d); the first reciprocal-lattice vector of the triangular lattice is at $G_1a_B^*=3.81=0.952\,(2k_Fa_B^*)$. Triangles indicate roton minima; imaginary frequencies are plotted as $-\sqrt{|\omega^2|}$ using dashed lines below zero.}
  \label{fig:roton_softening}
\end{figure}

To study the collective modes and instability of the exciton condensate we use the time-dependent Hartee-Fock(TDHF) theory. TDHF mode of momentum $\bq$ is a superposition of interband particle--hole fluctuations $\gamma^{\dagger}_{n,\bk+\bq}\gamma_{-m,\bk}$ and their conjugates, where $\gamma_{a,\bk}$ annihilates an electron in an eigenstate of the HFB Hamiltonian, and $-m$ ($n$) labels the occupied (empty) bands, $E_{-m,\bk}<0<E_{n,\bk}$. Each fluctuation moves an electron from an occupied to an empty band and conserves the total electron number in the bilayer system. Because these bands mix the layers, the fluctuations contain intralayer density components, $e^{\dagger}e$ and $h^{\dagger}h$, and interlayer components, $e^{\dagger}h^{\dagger}$ and $he$, which transfer an electron between the layers. The soft mode can therefore involve both density modulations and spatially varying interlayer coherence.

We first establish the uniform exciton condensate as a reference state at equal electron and hole masses, $R=m_h/m_e=1$. The matched normal-state Fermi contours favor a BCS-like instability in the interlayer particle--hole channel, characterized by $\Phi_{\bk}=\langle h_{-\bk}e_{\bk}\rangle\neq0$. At $d=3$~nm, the associated excitonic field $\Sigma^{X}_{\bk}$ opens a gap at the underlying Fermi surface [Fig.~\ref{fig:roton_softening}(a)]. The condensate breaks the $U(1)$ symmetry generated by $N_e+N_h$, giving rise to the Goldstone mode [Fig.~\ref{fig:roton_softening}(b)]. This mode disperses linearly at long wavelengths and remains below the particle--hole continuum over the plotted momentum range.

Mass asymmetry progressively destabilizes the uniform condensate.
At fixed $d=6$~nm, increasing $R$ lowers the Goldstone branch
over a broad momentum range and deepens a roton minimum near $2k_F$
[Fig.~\ref{fig:roton_softening}(c)]. The $R=20$ and $30$ curves exhibit
$\omega^2<0$ around the roton minimum, indicating a dynamical instability
at finite momentum.
The normal-state Fermi contours remain matched throughout
this scan, while the hole kinetic energy is suppressed when increasing $R$ at fixed $r_s^e=2$. Mass asymmetry thus enhances the
relative importance of intralayer interactions in the hole layer.

Increasing the interlayer separation at fixed $R=10$ also deepens the
roton minimum, with $\omega^2<0$ for $d=7.5$ and $8$~nm
[Fig.~\ref{fig:roton_softening}(d)].
In this scan, the long-wavelength dispersion changes little with $d$, whereas the mode near $2k_F$ softens strongly.
The finite-momentum instability therefore develops while the
long-wavelength mode remains stable.
Here, $r_s^e=2$ and $r_s^h=20$ are fixed: increasing $d$ reduces
$v_1(q)=v_0(q)e^{-qd}$ at finite momentum without changing the bare
intralayer interactions.
This behavior is consistent with competition between
interlayer coherence and spatial ordering in the strongly interacting
hole layer~\cite{BonitzEtAl2005,LudwigEtAl2007,SchleedeEtAl2012,BoeningFilinovBonitz2011,AroraMoudgil2021,NashabehFu2026}.
In the limit $d\rightarrow\infty$, the two layers are decoupled,
and a hole crystal coexisting with an electron metal (HC+EM) is a candidate state at the mean-field level.

The separation between the equilibrium crystallization boundary and the roton instability admits a natural analogy with a supercooled liquid. At the first-order transition, the dipolar Wigner crystal becomes energetically favorable while the uniform exciton condensate retains a finite roton gap and remains locally stable. The condensate can therefore persist as a metastable liquid beyond the equilibrium phase boundary, analogous to a liquid surviving below its freezing temperature. Physically, infinitesimal density modulations still cost energy even though a finite-amplitude crystalline state already has a lower total energy. This can be explained through Landau theory: a density wave on a triangular lattice admits a cubic term in the free energy, so that freezing is first order when the roton gap is still finite~\cite{AlexanderMcTague1978,PomeauRica1994,Nozieres2004}. Superfluid $^4$He is the classic example, its roton gap remain open under pressure before the liquid solidifies~\cite{DietrichEtAl1972}. A similar scenario has been analyzed for electron--hole bilayers, where a supersolid state is proposed to be candidate intermediate phase~\cite{JoglekarBalatskyDasSarma2006,Ye2008,FilShevchenko2020}. Upon further tuning the layer separation or mass ratio, the roton gap vanishes and the metastable condensate loses local stability. Within our HFB and TDHF description, the roton-instability line thus represents the spinodal limit of the uniform condensate, shown as the dashed line in Fig.~\ref{fig:mass_asymmetric_bilayer}(c). This picture provides a quantum analogue of supercooling, with the metastability controlled by interaction and kinetic-energy scales. We also note two competing phases that lie beyond our present mean-field framework at commensurate filling. First, an excitonic supersolid may emerge within the hatched metastable region if the unit-cell commensurability constraint is relaxed to accommodate incommensurate density waves near the $2k_F$ roton minimum. We examine this possibility below. Second, at large $R$ and intermediate $d$, the vicinity of the phase boundary may host an exciton Mott insulator, which is a correlated state of tightly bound electron--hole pairs that shares the spatial symmetry of the dipolar Wigner crystal but cannot be captured by our mean-field ansatz. The first-order EC--dWC transition we find should therefore be distinguished from a bosonic superfluid--Mott transition of tightly bound excitons.

\subsection{Phonon modes of the dipolar Wigner crystal}

\begin{figure}[t]
  \includegraphics[width=\linewidth]{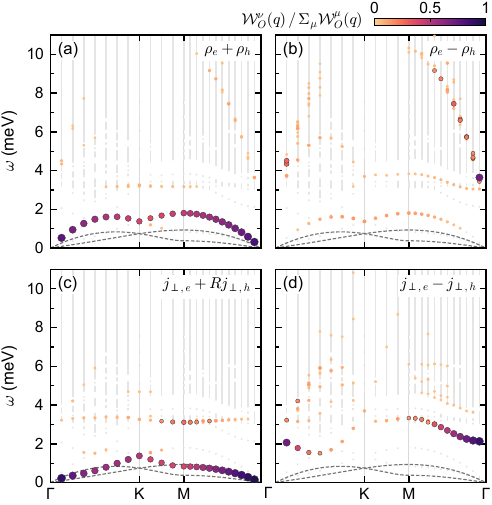}
  \caption{TDHF collective-mode spectrum of the dipolar Wigner crystal at $R=10$ and $d=8$~nm along $\Gamma$--K--M--$\Gamma$ ($24\times24$ mesh, $N_k=576$, two reciprocal-lattice shells; all $20\,736$ positive-frequency modes at each of the 23 wave vectors). All panels show the same excitation energies; marker color and area encode the normalized spectral residue $\mathcal{W}_O^{\nu}(\mathbf{q})/\sum_{\mu>0}\mathcal{W}_O^{\mu}(\mathbf{q})$ (Eq.~\eqref{eq:spectral_weight}) of mode $\nu$ for the operator $O$ indicated in each panel: (a) total density $\rho_e+\rho_h$ (longitudinal acoustic), (b) relative density $\rho_e-\rho_h$ (longitudinal optical), (c) transverse mass current $j_{\perp,e}+Rj_{\perp,h}$ (transverse acoustic), and (d) transverse charge current $j_{\perp,e}-j_{\perp,h}$ (transverse optical). Gray dots mark modes with normalized residue below $0.03$. Dashed curves are the acoustic phonons of the classical bilayer crystal; the classical optical branches lie at $\omega\simeq 19$--$20$~meV, above the displayed window.}\label{fig:dipolar_collective_modes}
\end{figure}

\begin{figure}[h]
  \includegraphics[width=\columnwidth]{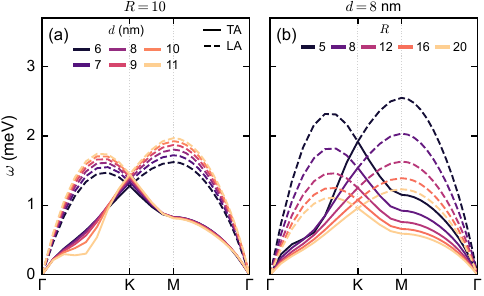}
  \caption{Acoustic phonon branches of the dipolar Wigner crystal along $\Gamma$--K--M--$\Gamma$, TA (solid) and LA (dashed), identified from the TDHF response as in Fig.~\ref{fig:dipolar_collective_modes} and followed by continuity from $\Gamma$; computed on a $24\times24$ mesh ($N_k=576$, 24 wave vectors on the path). (a)~$R=10$ for $d=6$--$11$~nm. (b)~$d=8$~nm for $R=5$--$20$. Other parameters as in Fig.~\ref{fig:dipolar_collective_modes}; the optical gap and the full $(R,d)$ dependence are summarized in Appendix~\ref{app:dwc-map}.}
  \label{fig:tdhf_cuts}
\end{figure}

We next examine the collective excitations of the dipolar Wigner crystal that competes with the uniform exciton condensate. Figure~\ref{fig:dipolar_collective_modes} shows a representative TDHF spectrum at mass ratio $R=10$ and interlayer separation $d=8$~nm. The spectrum contains acoustic modes describing collective lattice deformations and optical modes associated with relative motion of the electron and hole layers. Their dispersions allow us to assess the stability of the crystalline state and compare its dynamics with those of a classical bilayer crystal~\cite{KalmanEtAl2007,GoldenEtAl2012}.

To identify the character of the collective modes, we resolve the spectrum by its density and current responses. Within the quasi-boson approximation, the low-energy Hamiltonian reduces to a system of noninteracting bosons whose modes are the phonon-like excitations $|\nu,\mathbf q\rangle$ above the dWC HF ground state $|0\rangle$, with $\nu$ the mode index and $\mathbf q$ the crystal momentum~\cite{QBA_Tobi_Chunli, WC_Ambuj}. The spectral weight of mode $\nu$ for a probe operator $\hat O_{\mathbf q}$ is
\begin{equation}
    \mathcal{W}_O^\nu(\mathbf q)
    =
    \left|
    \langle \nu,\mathbf q|\hat O_{\mathbf q}|0\rangle
    \right|^2.
    \label{eq:spectral_weight}
\end{equation}
For the longitudinal response we use the density combinations $\rho_e+\rho_h$ and $\rho_e-\rho_h$, where $\rho_e$ and $\rho_h$ are the electron and hole number densities. For the transverse response we use $j_{\perp,e}+Rj_{\perp,h}$ and $j_{\perp,e}-j_{\perp,h}$, which are proportional to the mass and charge currents, respectively. Here $j_{\perp,\alpha}$ is the particle current in layer $\alpha\in\{e,h\}$ perpendicular to $\mathbf q$, and $R=m_h/m_e$. Each panel of Fig.~\ref{fig:dipolar_collective_modes} shows the same excitation energies, colored by the normalized weight $\mathcal{W}_O^\nu(\mathbf q)/\sum_{\mu>0}\mathcal{W}_O^\mu(\mathbf q)$, where the sum runs over all positive-frequency modes at the same $\mathbf q$. The operator definitions and the TDHF evaluation of $\mathcal{W}_O^\nu$ are given in Appendix~\ref{app:response}.

The density-sum and transverse mass-current responses identify longitudinal and transverse acoustic branches associated with broken translational symmetry. Higher-energy optical excitations carry substantial relative-density and relative-current weight, consistent with relative motion of the two layers. As shown in Fig.~\ref{fig:dipolar_collective_modes}, no finite-momentum dynamical instability is found along the sampled path, supporting the local stability of the competing crystal. The same holds across the dipolar-crystal region of Fig.~\ref{fig:mass_asymmetric_bilayer}(c) except near its boundary with the HC+EM phase, where the TA branch near $\Gamma$K$/2$ becomes unstable. The TDHF spectra of all ground-state cells are summarized in Appendix~\ref{app:dwc-map}.  Comparison with the classical bilayer phonons~\cite{KalmanEtAl2007,GoldenEtAl2012}, shown by dashed curves, reveals substantial changes in the excitation energies. The longitudinal acoustic branch lies appreciably above its classical counterpart over much of the Brillouin zone, while the transverse branch exhibits a more momentum-dependent deviation. Conversely, the optical excitations are strongly softened relative to the classical prediction, whose optical branches lie above the displayed energy window. Relative-layer motion therefore occurs much closer in energy to the acoustic sector in the TDHF crystal. These contrasting shifts demonstrate that the quantum spectrum cannot be obtained by uniformly rescaling the classical phonon energies. They reflect the self-consistent electronic structure, including exchange and the spatial extent and rearrangement of the carrier densities, although the dispersion alone does not isolate these contributions.

The acoustic and optical modes respond differently to mass asymmetry and
layer separation. Increasing $R$ lowers both acoustic branches
[Fig.~\ref{fig:tdhf_cuts}(b)], with small-wave-vector velocity estimates approximately
following the $(m_e+m_h)^{-1/2}$ scaling expected for the common motion of
the two layers. The optical gap instead
is much more sensitive to $d$ (Appendix~\ref{app:dwc-map}). Increasing the separation
weakens the interlayer restoring force and reduces the induced electron
density modulation, softening the relative-layer oscillation. The
resulting optical frequencies span $0.3$--$1.2$~THz.
The longitudinal mode compresses both layers in phase. The electrostatic
restoring force contributes
$c_{L,\mathrm H}^{\,2} = ne^2d/\varepsilon_0\varepsilon_r(m_e+m_h)
$
to the squared sound velocity. This contribution increases with $d$,
consistent with the TDHF trend.
Meanwhile the transverse velocity decreases weakly. At finite momentum,
the softened optical mode hybridizes with the TA branch, producing the
shoulder in Fig.~\ref{fig:tdhf_cuts}(a) at larger separations. Thus the evolution of the
acoustic spectrum reflects both changes in elastic stiffness and coupling
to relative-layer motion. No additional finite-momentum soft mode is
resolved near the condensate boundary, consistent with the first-order
transition inferred from the ground-state energies. Near the opposite boundary, this hybridized TA branch near $\Gamma$K$/2$ softens to zero, so that the dWC becomes dynamically unstable while its electron gap is still open (Appendix~\ref{app:dwc-map}).

\begin{figure}
    \centering
    \includegraphics[width=\columnwidth]{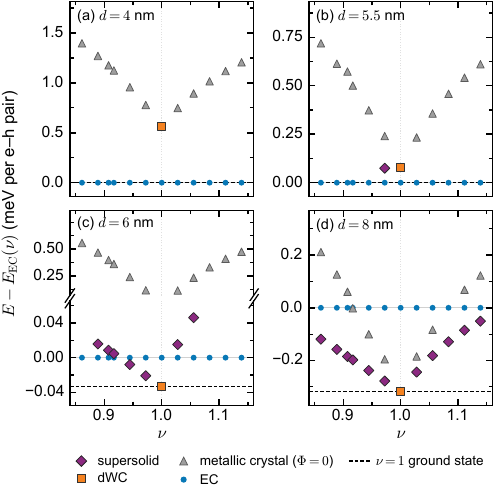}
    \caption{Energy of self-consistent solutions versus filling $\nu$ at $R=10$ for (a) $d=4$, (b) $5.5$, (c) $6$, and (d) $8$~nm, measured from the uniform condensate at the same $\nu$, which removes the $\nu$-dependent finite-size offset. The $d^* \approx 5.9~nm$ at $R = 10$. Symbols mark the distinct solutions obtained from all seed states considered (see main text) at each $\nu$, labeled by the criteria of Table~\ref{tab:criteria}. The dashed line is the energy of the $\nu=1$ ground state. Supersolid solutions appear at $\nu\neq1$ for $d\gtrsim5.5$~nm, with coherence growing with $|1-\nu|$, but at every $d$ they lie above the lowest $\nu=1$ state. $N_k=324$, $N_G=19$.}
    \label{fig:ene_vs_nu}
\end{figure}

\subsection{Possible supersolid phase}

As we mentioned above, the broad metastable regime of the exciton condensate naturally suggests a potential intermediate supersolid phase, characterized by the coexistence of interlayer coherence and broken translational symmetry. Within our HFB variational space, such states are directly captured by the simultaneous emergence of a nonzero anomalous pairing $\Phi$ and finite density-wave harmonics $\tilde{n}_{\alpha}(\mathbf{G}\neq0)$. Notice that the phase diagram in Fig.~\ref{fig:mass_asymmetric_bilayer}(c) is restricted to commensurate crystals with one electron and one hole per unit cell ($\nu=1$), which rigidly fixes the fundamental reciprocal lattice vector $G$. Since the roton minimum lies close to, but does not exactly coincide with $G$, relaxing this commensurability constraint could render a supersolid with a slightly shifted lattice periodicity energetically competitive.

To explore this possibility, we relax the $\nu=1$ constraint at fixed total electron and hole densities, thereby tuning the fundamental reciprocal-lattice vector via $G\propto\nu^{-1/2}$. Figure~\ref{fig:ene_vs_nu} displays the energies of the resulting self-consistent solutions as a function of $\nu$ at $R=10$ for several interlayer separations. For $d\gtrsim5.5$~nm, self-consistent supersolid solutions indeed emerge at $\nu\neq1$. Notably, the interlayer coherence grows as the filling is detuned from unity, demonstrating that lifting the commensurability constraint enables the crystalline state to partially restore condensate order. This behavior can be viewed as a fermionic analog of the Andreev--Lifshitz--Chester defect-condensation mechanism~\cite{AndreevLifshitz1969,Chester1970,ContiEtAl2023Chester}: detuning $\nu$ from unity at fixed density dopes the insulating crystal with itinerant vacancies ($\nu<1$) or interstitials ($\nu>1$). These defects are fermions that partially fill the crystal bands of the two layers, and they condense through interlayer pairing. The defect formation energy here remains slightly positive at the HFB level, so the supersolid is not ground state in our calculation. Nevertheless, this near-degeneracy establishes the supersolid as an important low-lying competing phase near the first order EC--dWC transition, motivating future investigations that incorporate quantum fluctuations.

\section{Discussion}\label{sec:discussion}

Experimental realizations of mass-asymmetric electron--hole bilayers fall into two main categories. The first comprises transition-metal dichalcogenide (TMD) systems, including bilayer graphene (BLG)/TMD~\cite{ZhangEtAl2024RemoteCoulomb} and $\text{TMD}_1/\text{TMD}_2$ heterostructures~\cite{WangEtAl2019Nature,MaEtAl2021Nature,NguyenEtAl2025Science,QiEtAl2025Science}, with recent evidence for exciton crystallization~\cite{QiEtAl2026ExcitonCrystal}. The second involves interfacial charge-transfer heterostructures, such as (multilayer) graphene/Mott insulator (MI) assemblies~\cite{RizzoEtAl2020,WangEtAl2020ModulationDoping,WangEtAl2022NatNano,LuEtAl2023NatCommun,TilakEtAl2024NatCommun,YangEtAl2025NSR,KimEtAl2026NatCommun,ShiEtAl2026ChargeTransfer,MiaoEtAl2025}. Here, work-function mismatch drives spontaneous charge transfer, forming spatially separated electron--hole layers with strong mass asymmetry inherited from the heavy MI quasiparticles, while a variable-thickness hBN spacer tunes the carrier density and interlayer spacing. Using BLG ($m^* \approx 0.03\,m_0$~\cite{ZouHongZhu2011}) at a typical density $n \sim 10^{11}$~cm$^{-2}$ gives $r_s \approx 2$, directly within our calculated regime.

Discriminating between a uniform exciton condensate, an exciton supersolid, and an incoherent dipolar Wigner crystal in a real device requires complementary probes of transport and translational order. Optical exciton spectroscopy and scanning tunneling microscopy have revealed electronic crystallization in two-dimensional materials~\cite{SmolenskiEtAl2021,TsuiEtAl2024}, providing possible routes to probing density order in bilayers. To test phase coherence, lateral Josephson junctions have been proposed for electron--hole bilayers, where exciton transport across a weak link depends on the phase difference between the two superfluid regions~\cite{PascucciEtAl2022}. Terahertz absorption offers a contact-free probe of the incoherent crystal. A uniform in-plane field drives only the relative motion of the two layers, which in the dWC is a gapped transverse optical phonon in the terahertz range, whereas the uniform condensate has no such mode. Combining phase-sensitive measurements with evidence for spontaneous density order, such as this phonon line, in the same phase would offer a route to distinguishing excitonic supersolidity from an incoherent crystal.

We also note that the exotic phases near the phase transition can be more fruitful once band geometry is taken into consideration, which we leave for an upcoming work.

\begin{acknowledgments}
We thank Xi Dai, Yongxin Zeng, Yuelin Shao for fruitful discussion and Chunli Huang for previous collaboration. W. Miao acknowledges a fellowship from the Center for Theory of Emergent Quantum Matter (C-TEQ) at Penn State. W. Miao also thanks the Aspen Center for Physics, where this work was partly done, which is supported by a grant from the Simons Foundation (1161654, Troyer), and the Kavli Institute for Theoretical Physics (KITP), supported in part by the National Science Foundation under Grant No. PHY-2309135. A. Jain is grateful to the University of Kentucky Center for Computational Sciences and Information Technology Services Research Computing for their support and use of the Morgan Compute Cluster and associated research computing resources.

\end{acknowledgments}

\section*{Declaration of AI usage}
We used Claude Opus 5.5 (Anthropic) and ChatGPT 6 Astra (OpenAI) in parts of the numerical calculations, plotting, and polishing of the manuscript. The numerical results were checked by the authors independently, and the AI models were mainly used to optimize the existing code for TDHF we developed in Refs.~\cite{WC_Ambuj,MiaoChenYanHuang2026}.

\bibliography{reference}

\onecolumngrid

\appendix

\section{Details of mean-field numerical calculations}
\label{app:hf}

\subsection{HFB Hamiltonian}

In the Nambu basis of Eq.~\eqref{eq:nambu}, the HFB Hamiltonian at each $\bk$ in the Brillouin zone (BZ) is the $2N_G\times2N_G$ matrix
\begin{equation}
  h(\bk)=
  \begin{pmatrix}
    T_e(\bk)+\Sigma^{H}_{e}+\Sigma^{F}_{e}(\bk)-\mu & \Sigma^{X}(\bk)\\[2pt]
    \Sigma^{X\dagger}(\bk) & \mu-\big[T_h(-\bk)+\Sigma^{H}_{h}+\Sigma^{F}_{h}(-\bk)\big]^{*}
  \end{pmatrix},
  \label{eq:app-hHF}
\end{equation}
with $[T_{\alpha}(\bk)]_{\bG\bG'}=\delta_{\bG\bG'}\hbar^2|\bk+\bG|^2/2m_\alpha$ for $\alpha=e,h$. The lower block acts on $h^\dagger_{-\bk+\bG}$, so it is minus the complex conjugate of the hole Hamiltonian at $-\bk$. Since $N_e-N_h$ is conserved and fixed to zero below, $\mu_e-\mu_h$ only shifts all eigenvalues of $h(\bk)$ uniformly, and only $\mu=(\mu_e+\mu_h)/2$ enters.

With $v_{ee}=v_{hh}=v_0$ and $v_{eh}=v_{he}=-v_1$, the Hartree field is
\begin{equation}
  [\Sigma^{H}_{\alpha}]_{\bG\bG'}
  =\frac{N_k}{S}\,(1-\delta_{\bG\bG'})\sum_{\beta=e,h}v_{\alpha\beta}(|\bG-\bG'|)\,n_{\beta}(\bG-\bG').
  \label{eq:app-hartree}
\end{equation}
All exchange fields follow from
\begin{equation}
  \big[\mathcal{X}_{v}[M](\bk)\big]_{\bG\bG'}
  =-\frac{1}{S}\sum_{\bk'\in{\rm BZ}}\sum_{\bG''}
  v(|\bk-\bk'-\bG''|)\,M_{\bG+\bG'',\,\bG'+\bG''}(\bk'),
  \label{eq:app-X}
\end{equation}
as $\Sigma^{F}_{\alpha}=\mathcal{X}_{v_0}[\rho^{\alpha}]$ for $\alpha=e,h$, with $\rho^{h}(-\bk)=[\bar\rho^{h}(\bk)]^{*}$ read off the lower block of $\Gamma(\bk)$. Because the hole slot carries $-\bk+\bG$, the two indices of $\Phi$ shift oppositely in the pairing field,
\begin{equation}
  [\Sigma^{X}(\bk)]_{\bG\bG'}
  =-\frac{1}{S}\sum_{\bk'\in{\rm BZ}}\sum_{\bG''}
  v_1(|\bk-\bk'-\bG''|)\,\Phi_{\bG+\bG'',\,\bG'-\bG''}(\bk').
  \label{eq:app-SX}
\end{equation}
Because Eq.~\eqref{eq:Hamiltonian} is normal ordered with respect to the hole vacuum, no filled-band contribution appears. The pairing field $\Sigma^{X}$ is generated by the interlayer attraction; its phase is not fixed by $H$, reflecting the broken $U(1)$ symmetry generated by $N_e+N_h$.

\subsection{Self-consistency and energy}

We diagonalize $h(\bk)$ at every $\bk$ and occupy the lowest $N_kN_G$ of the $2N_kN_G$ eigenvalues, which defines $\Gamma(\bk)$. Since $\mathrm{Tr}\,\Gamma(\bk)=N_G+\mathrm{Tr}\,\rho^{e}(\bk)-\mathrm{Tr}\,\rho^{h}(-\bk)$, this half filling gives $N_e=N_h$ independently of $\mu$, while the number of occupied states at a given $\bk$ may differ from $N_G$. Within each iteration, $\mu$ is fixed by bisection so that $N_e+N_h=2N_k$ and $n_e=n_h=n$.

The energy of the converged state is $E=E^{\rm kin}+E_{\rm int}$, with $E_{\rm int}=E^{H}+E^{F}+E^{X}$ given by Eqs.~\eqref{eq:energy} and
\begin{equation}
  E^{\rm kin}=\sum_{\bk,\alpha}\mathrm{Tr}\big[T_\alpha(\bk)\,\rho^{\alpha}(\bk)\big].
  \label{eq:app-E}
\end{equation}
The coherence part is
\begin{equation}
  E^{X}=\mathrm{Re}\sum_{\bk}\mathrm{Tr}\big[\Sigma^{X}(\bk)\,\Phi^{\dagger}(\bk)\big]\le0 ,
  \label{eq:app-EX}
\end{equation}
nonpositive because $v_1(q)>0$. We characterize converged states by $E^{X}/E_{\rm int}$ and by the first-shell average $\langle|n_\alpha(\bG_1)|\rangle=\tfrac16\sum_{|\bG|=G_1}|n_\alpha(\bG)|$.

\subsection{Numerical details}

Momenta are measured in units of $G_1=4\pi/(\sqrt3\,a)$ and energies in ${\rm Ry}_e=\hbar^2/[2m_e(a_B^e)^2]$. The problem then depends only on $r_s^e$, $R=m_h/m_e$, and $G_1d$. We use an $N_k=324$ mesh and the $N_G=19$ shortest reciprocal-lattice vectors.

The integrable Coulomb singularity of Eq.~\eqref{eq:app-X} at $\bk'+\bG''=\bk$ is omitted. Since $v_1\to v_0$ as $q\to0$, the omitted term in $h(\bk)$ is proportional to $\Gamma(\bk)$, up to a constant in the hole block. Because $\Gamma$ is idempotent and $N_h$ is fixed, its contribution to the energy is state independent to leading order and cancels in energy differences.

Each $(R,d)$ point is converged from the four classes of initial states below, and the lowest-energy solution is retained:
\begin{center}
\begin{tabular}{lcc}
\hline\hline
Initial state & $n_\alpha(\bG\neq0)$ & $\Phi$ \\
\hline
uniform condensate & $0$ & finite \\
aligned dipolar crystal & both layers & $0$ \\
hole crystal, uniform electrons & hole layer only & $0$ \\
coexistence & both layers & finite \\
\hline\hline
\end{tabular}
\end{center}
The coexistence seed is required because the self-consistency map preserves the symmetry of the initial state. A seed with $\Phi=0$ gives $\Sigma^{X}=0$ at every iteration, and a seed with $n_\alpha(\bG\neq0)=0$ keeps $h(\bk)$ diagonal in $\bG$, so neither can develop the other order.

\subsection{Order parameters and layer quasiparticle bands at commensurate filling}

\begin{figure}[t]
  \includegraphics[width=\linewidth]{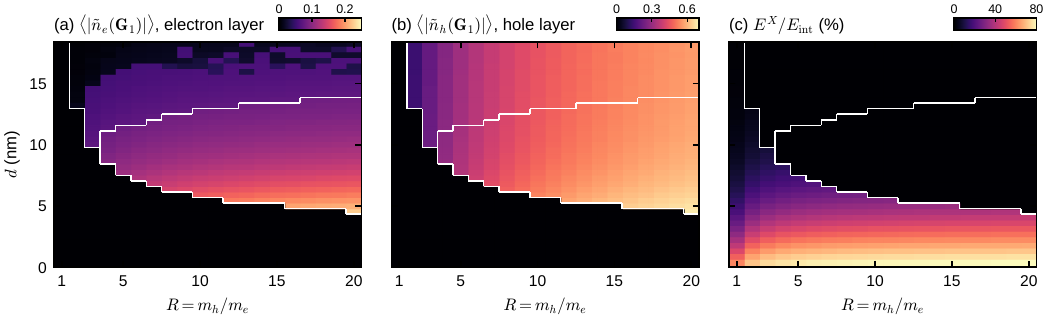}
    \caption{Mean-field ground state at $r_s^e = 2$ in the $(R, d)$ plane, $R = m_h/m_e$. (a), (b) First-shell density Fourier amplitude $\langle|\tilde{n}_\alpha(\bG_1)|\rangle$, with $\alpha \in \{e, h\}$. (c) Percentage of the interaction energy carried by interlayer coherence. White lines are the phase boundaries of Fig.~\ref{fig:mass_asymmetric_bilayer}(c). Coherence occupies small $d$, where both layers are uniform ($\langle|\tilde{n}_{e,h}(\bG_1)|\rangle < 10^{-3}$). For $R \geq 2$ the hole layer crystallizes at large $d$, while the electron layer stays gapless with weak modulation $\langle|\tilde{n}_e(\bG_1)|\rangle \lesssim 0.07$ (HC+EM). For $R \geq 4$ both layers are gapped in an intermediate $d$ window (dWC, $\langle|\tilde{n}_e(\bG_1)|\rangle = 0.08$--$0.24$), which widens with $R$. Coherence and crystalline order never coexist.}
  \label{fig:order_parameter}
\end{figure}

Figure~\ref{fig:order_parameter} shows the density order in each layer, and  coherence over scanned parameter space(see Figure~\ref{fig:mass_asymmetric_bilayer}(c)). The coherence fraction $E^{X}/E_{\rm int}$ is finite only in the condensate, where both layers are uniform to $\langle|\tilde n_\alpha(\bG_1)|\rangle<10^{-2}$, while every crystalline state has $E^{X}/E_{\rm int}<10^{-3}$. No converged state carries both density and coherence orders, including those started from the coexistence seed. In the crystalline region the hole amplitude, $\langle|\tilde n_h(\bG_1)|\rangle=0.14$--$0.67$, exceeds the electron amplitude everywhere. The heavy holes form the lattice, and the electron modulation, $\langle|\tilde n_e(\bG_1)|\rangle\le0.22$, is induced by it through the interlayer Hartree term, $v_1(G_1)\propto e^{-G_1d}$, so it weakens as $d$ grows.

\begin{figure}[h]
  \includegraphics[width=\linewidth]{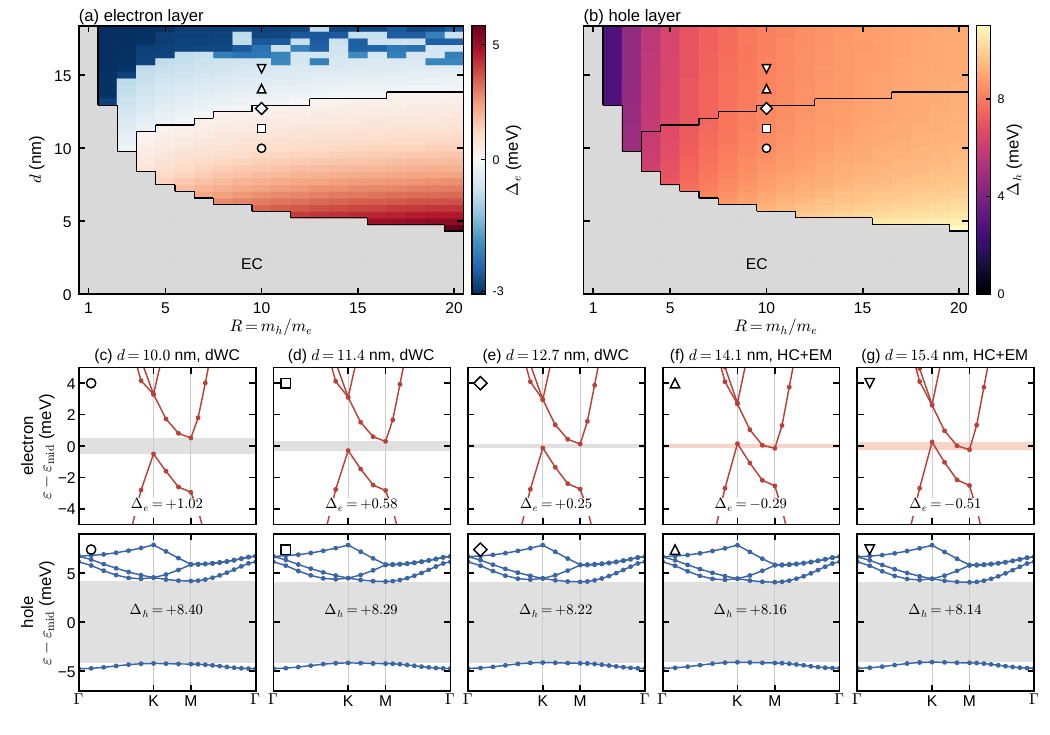}
    \caption{Layer quasiparticle gaps and bands of the incoherent ($\Phi=0$) states at $r_s^e=2$. (a), (b) Indirect gaps $\Delta_e$ and $\Delta_h$ in the $(R,d)$ plane; the condensate region is gray and black lines are the phase boundaries of Fig.~\ref{fig:mass_asymmetric_bilayer}(c). (c)--(g) Electron (top) and hole (bottom) bands along $\Gamma$--K--M--$\Gamma$ at the points marked by the same symbols in (a), (b), at $R=10$. Energies are measured from each layer's chemical potential; gray shading marks a gap and red shading a band overlap ($\Delta<0$). Across the dWC--HC+EM boundary the electron gap closes at M, while the hole layer stays insulating with $\Delta_h\simeq 8$~meV. $N_k=324$, $N_G=19$.}
  \label{fig:band_st_and_gap}
\end{figure}

The dWC and HC+EM phases are distinguished by the electron layer quasiparticle band gap [Figure~\ref{fig:band_st_and_gap}(a)]. Across the crystalline region $\Delta_e$ decreases continuously with $d$ and changes sign at the boundary between these phases. The smallest gap in the dWC is $0.17$~meV and the largest gap in HC+EM is $-0.04$~meV, so the classification does not depend on the choice of $\delta$. The bands at $R=10$ show how the gap closes. In the dWC the lowest quasiparticle electron band has its maximum at K and the second band its minimum at M [Figure~\ref{fig:band_st_and_gap}(c)--(e)]. As $d$ increases the two edges approach and overlap beyond $d\simeq13$~nm, leaving electron quasiparticle pockets at M and hole pockets at K [Figure~\ref{fig:band_st_and_gap}(f), (g)]. Because the electron layer holds exactly one electron per unit cell, the overlap produces a compensated semimetal rather than a partially filled band. The hole layer stays insulating, with $\Delta_h=2.7$--$11$~meV in the crystalline region and a nearly flat occupied band, reflecting the strongly localized heavy holes [Figure~\ref{fig:band_st_and_gap}(b)].

\section{Response functions and spectral weights}
\label{app:response}

\subsection{Layer operators}

Let $\alpha^{\dagger}$, with $\alpha=e,h$, be the creation operator in the electron or hole layer, with $\mathbf{k}$ in the first Brillouin zone and $\mathbf{G}$ a reciprocal lattice vector. The density and paramagnetic particle current of layer $\alpha$ are
\begin{align}
  \rho_{\alpha}(\mathbf{q}) &= \sum_{\mathbf{k},\mathbf{G}}
    \alpha^{\dagger}_{\mathbf{k}+\mathbf{q}+\mathbf{G}}\alpha^{\phantom{\dagger}}_{\mathbf{k}+\mathbf{G}},
  \label{eq:app-rho}\\
  \mathbf{j}_{\alpha}(\mathbf{q}) &= \frac{1}{m_{\alpha}}\sum_{\mathbf{k},\mathbf{G}}
    \Big(\mathbf{k}+\mathbf{G}+\tfrac{\mathbf{q}}{2}\Big)\,
    \alpha^{\dagger}_{\mathbf{k}+\mathbf{q}+\mathbf{G}}\alpha^{\phantom{\dagger}}_{\mathbf{k}+\mathbf{G}},
  \label{eq:app-current}
\end{align}
with longitudinal and transverse components
\begin{equation}
  j_{\parallel,\alpha}(\mathbf{q}) = \hat{\mathbf{q}}\cdot\mathbf{j}_{\alpha}(\mathbf{q}),
  \qquad
  j_{\perp,\alpha}(\mathbf{q}) = (\hat{\mathbf{z}}\times\hat{\mathbf{q}})\cdot\mathbf{j}_{\alpha}(\mathbf{q}).
  \label{eq:app-jparperp}
\end{equation}
Without interlayer tunneling, $\rho_{\alpha}(0)$ are separately conserved, and translation invariance conserves $\mathbf{P}=\sum_{\alpha}m_{\alpha}\mathbf{j}_{\alpha}(0)$. The charge current $\mathbf{j}_{e}(0)-\mathbf{j}_{h}(0)$ is relaxed by the interlayer interaction and is therefore not conserved. It determines the $\mathbf{q}=0$ optical response.

\subsection{Response functions and spectral weight}

The TDHF modes $\nu$ at momentum $\mathbf{q}$ carry amplitudes $(X^{\nu},Y^{\nu})$ on the occupied--empty and empty--occupied pairs of the HF ground state, normalized in the metric $\mathcal{I}=\mathrm{diag}(\mathbb{I},-\mathbb{I})$ so that $\sum_{ph}(|X^{\nu}_{ph}|^{2}-|Y^{\nu}_{ph}|^{2})=\mathcal{N}_{\nu}=\pm1$, with $+1$ on the branch $\omega_{\nu}>0$ and $-1$ on the branch $\omega_{\nu}<0$. The transition amplitude of an operator $O$ carrying momentum $\mathbf{q}$ is
\begin{equation}
  a^{\nu}_{O}(\mathbf{q}) \;=\; \sum_{ij}O^{*}_{ij}X^{\nu}_{ij}(\mathbf{q})
  \;=\; \sum_{n\neq 0,\mathbf{k}}\Big[
    O^{*}_{n\mathbf{k}+\mathbf{q},\,0\mathbf{k}}\,X^{\nu}_{n\mathbf{k}}(\mathbf{q})
    + O^{*}_{0\mathbf{k},\,n\mathbf{k}-\mathbf{q}}\,Y^{\nu}_{n\mathbf{k}}(-\mathbf{q})\Big],
  \label{eq:app-amp}
\end{equation}
with $0$ the occupied HF bands and $n\neq0$ the empty ones, and the spectral weight (residue) of the mode is
\begin{equation}
  \mathcal{W}^{\nu}_{O}(\mathbf{q}) \;=\; \big|a^{\nu}_{O}(\mathbf{q})\big|^{2}.
  \label{eq:app-residue}
\end{equation}
The matrix elements $O_{ij}$ must be evaluated in the same HF orbitals in which the eigenvectors are expressed: a degenerate HF band may be returned in a rotated gauge, which leaves the band energies unchanged but spoils Eq.~(\ref{eq:app-amp}).

In terms of these amplitudes the retarded response of $O_{2}$ to a field coupling to $O_{1}$ is
\begin{equation}
  \chi_{O_{2}O_{1}}(\mathbf{q},\Omega) \;=\; \sum_{\nu>0}\left[
    \frac{a^{\nu}_{O_{2}}a^{\nu*}_{O_{1}}}{\Omega-\omega_{\nu}+i0^{+}}
    \;-\; \frac{b^{\nu}_{O_{2}}b^{\nu*}_{O_{1}}}{\Omega+\omega_{\nu}+i0^{+}}\right],
  \qquad
  b^{\nu}_{O} \;=\; \sum_{ij}O^{*}_{ij}X^{\nu*}_{ji}.
  \label{eq:app-chi}
\end{equation}
The two numerators are distinct. Since $O$ carries momentum $\mathbf{q}$, the transposed amplitude $b^{\nu}_{O}$ involves the modes at $-\mathbf{q}$; the partner of $(X^{\nu},Y^{\nu})$ there, with eigenvalue $-\omega_{\nu}$, is $(Y^{\nu*},X^{\nu*})$, so that $X^{\bar{\nu}}_{ij}=X^{\nu*}_{ji}$ and
\begin{equation}
  b^{\nu}_{O}(\mathbf{q}) \;=\; a^{\bar{\nu}}_{O}(\mathbf{q}),
  \qquad \omega_{\bar{\nu}}=-\omega_{\nu}.
  \label{eq:app-b-identity}
\end{equation}
The second numerator is therefore the same contraction (\ref{eq:app-amp}) evaluated on the $\omega<0$ eigenvectors at the \emph{same} $\mathbf{q}$, and Eq.~(\ref{eq:app-chi}) collapses to a single sum over both branches,
\begin{equation}
   \chi_{O_{2}O_{1}}(\mathbf{q},\Omega) \;=\; \sum_{\nu}\,\mathcal{N}_{\nu}\,
   \frac{a^{\nu}_{O_{2}}a^{\nu*}_{O_{1}}}{\Omega-\omega_{\nu}+i0^{+}} ,
  \label{eq:app-chi-compact}
\end{equation}
Time-reversal symmetry relates the two branches, $b^{\nu}_{O_{2}}b^{\nu*}_{O_{1}}=s_{1}s_{2}\,a^{\nu}_{O_{2}}a^{\nu*}_{O_{1}}$ with $s=+1$ for $\rho$ and $s=-1$ for $\mathbf{j}$, so that the numerators coincide in the $\rho\rho$ and $jj$ channels and differ in sign in the mixed ones; we evaluate both branches explicitly and use this only as a check.

For $\Omega>0$ only the first pole contributes, and broadening the $\delta$ functions into Lorentzians of half-width $\eta$ gives the spectral function
\begin{equation}
  -\,\mathrm{Im}\,\chi_{O_{2}O_{1}}(\mathbf{q},\Omega)
  \;=\; \sum_{\nu>0}a^{\nu}_{O_{2}}a^{\nu*}_{O_{1}}\,
  \frac{\eta}{(\Omega-\omega_{\nu})^{2}+\eta^{2}} ,
  \label{eq:app-spectral}
\end{equation}
which for $O_{2}=O_{1}$ reduces to $\sum_{\nu>0}\mathcal{W}^{\nu}_{O}$ weighted by Lorentzians and is non-negative.

\subsection{Auto-responses}

We evaluate Eq.~(\ref{eq:app-spectral}) for the self responses $O_{2}=O_{1}=O$ of the four combinations that separate the branches of the crystal:
\begin{center}
\begin{tabular}{llll}
\hline\hline
Channel & Operator $O$ & Branch probed & Weight at $\mathbf{q}=0$ \\
\hline
in-phase density & $\rho_{e}+\rho_{h}$ & longitudinal acoustic (LA) & none \\
charge density & $\rho_{e}-\,\rho_{h}$ & longitudinal optical (LO) & none \\
mass current & $m_{e}j_{\perp,e}+m_{h}j_{\perp,h}$ & transverse acoustic (TA) & sliding mode only \\
charge current & $j_{\perp,e}-\,j_{\perp,h}$ & transverse optical (TO) & finite \\
\hline\hline
\end{tabular}
\end{center}
The in-phase combinations follow the center-of-mass motion of the two lattices and select the acoustic branches, while the charge combinations follow their relative motion and select the optical ones; densities probe longitudinal modes and $j_{\perp}$ the transverse ones. The first three operators are conserved at $\mathbf{q}=0$ and therefore carry no finite-frequency weight there, the mass current retaining only the zero-frequency sliding mode. The charge current is not conserved, and its $\mathbf{q}=0$ response gives the optical conductivity
\begin{equation}
  \mathrm{Re}\,\sigma(\omega)\;=\; \frac{\pi e^2}{S\omega}
    \sum_{\nu>0}\mathcal{W}^{\nu}_{J}\,\delta(\omega-\omega_{\nu}),
    \qquad \omega>0,\qquad J=\lim_{\mathbf{q}\to0}\big[j_{\perp,e}(\mathbf{q})-j_{\perp,h}(\mathbf{q})\big],
    \label{eq:app-sigma}
\end{equation}
whose weight is concentrated in the transverse optical mode. Cross responses $O_{2}\neq O_{1}$ follow from the same amplitudes; unlike the auto-responses they are not sign definite, and the sign of $-\mathrm{Im}\,\chi_{O_{2}O_{1}}$ at a mode records whether the two layers move in phase or out of phase in it.

The broadening $\eta$ in Eq.~(\ref{eq:app-spectral}) is a plotting parameter, not a lifetime: the TDHF eigenvalues are real, and a collective mode degenerate with the continuum of HF interband excitations shows up not as a width but as a fragmentation of its residue over the neighboring continuum modes, the finite-$N_{k}$ image of Landau damping.

\section{Collective modes across the dipolar-crystal region}
\label{app:dwc-map}
\begin{figure}[t]
  \includegraphics[width=\linewidth]{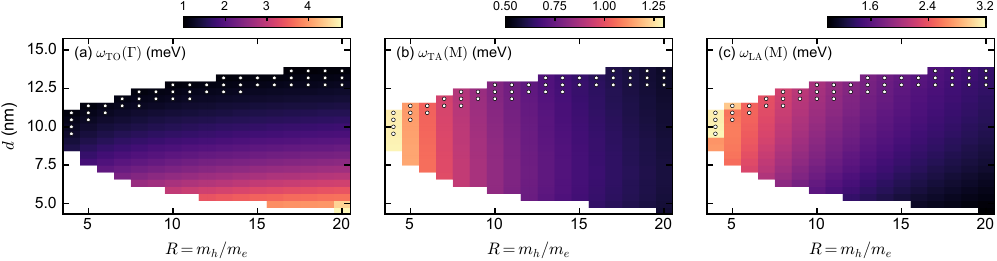}
  \caption{
  Phonon energies across the dipolar-crystal region of Fig.~\ref{fig:mass_asymmetric_bilayer}(c), for
all $268$ cells of the dWC region ($R=4$--$20$, $d=4.5$--$13.6$~nm in steps of $20/44$~nm). Here $n=10^{11}$~cm$^{-2}$, $m_e=0.07\,m_0$,
$\varepsilon_r=11.8$, $N_k=324$, and $N_G=19$.
(a) Optical gap at $\Gamma$, where the longitudinal and transverse
optical modes are degenerate. (b),(c) TA and LA energies at M.
Longitudinal modes are assigned by their density residues and
transverse modes by their rigid-displacement projections, as
described in this appendix. Each panel has its own color scale;
white cells lie outside the dipolar-crystal region. Dots mark cells in which the TA branch has imaginary frequency at $q=\Gamma\mathrm{K}/2$ (and in most of them also at $\Gamma\mathrm{K}/3$).}
  \label{fig:tdhf_dwc_map}
\end{figure}

We calculate the TDHF spectrum in all $268$ cells in which the dipolar
crystal is the HFB ground state in
Fig.~\ref{fig:mass_asymmetric_bilayer}(c). Each crystalline state is reconverged using an $18\times18$
momentum mesh and $N_G=19$ plane waves. At each of the $18$
independent wave vectors along $\Gamma$--K--M--$\Gamma$, we obtain
the $80$ eigenvalues closest to zero and their eigenvectors. The two
translational Goldstone modes are recovered at $\Gamma$. In $46$ cells near the boundary of the HC+EM phase, the TA branch has imaginary frequency at $q=\Gamma\mathrm{K}/2$, and in $26$ of them also at $\Gamma\mathrm{K}/3$, with $|\mathrm{Im}\,\omega|\le0.19$~meV. All other cells are stable along the sampled path.

The branches are resolved by their density and displacement character.
The longitudinal channels are the spectral residues of $\rho_e+\rho_h$
and $\rho_e-\rho_h$ defined in Appendix~\ref{app:response}. For the transverse modes in
the parameter scan, rigid layer displacements $\mathbf u_e$ and
$\mathbf u_h$ are fitted to the microscopic density fluctuations.
The projections $|u_{e\perp}+R u_{h\perp}|^2$ and
$|u_{e\perp}-u_{h\perp}|^2$ characterize common and relative motion,
respectively; these displacement projections specify polarization rather
than current spectral weight. Figure~\ref{fig:tdhf_dwc_map} assigns modes by their dominant
channel, while the acoustic branches in Fig.~\ref{fig:tdhf_cuts} are followed continuously
from $\Gamma$ through their density and displacement patterns. At
$\Gamma$, the degenerate optical modes are identified by their
relative-layer displacement.

Figure~\ref{fig:tdhf_dwc_map}(a) shows that the optical gap decreases monotonically with
$d$. Its scale follows from the interlayer Hartree
energy of a rigid relative displacement $\mathbf u$. For the aligned
triangular crystal, the energy change per electron--hole pair is
\begin{equation}
 \delta E_{eh}^{\mathrm H}
 =\frac{1}{2}K_{\mathrm{rel}}|\mathbf u|^2,
 \qquad
 K_{\mathrm{rel}}=
 \frac{n}{2}\sum_{\mathbf G\ne0}G^2v_1(G)
 \operatorname{Re}\!\left[n_e^*(\mathbf G)n_h(\mathbf G)\right].
 \label{eq:dwc_relative_stiffness}
\end{equation}
The corresponding optical energy is
\begin{equation}
 \Delta_{\mathrm{opt}}^{\mathrm{rig}}
 =\hbar\sqrt{K_{\mathrm{rel}}/\mu_{\mathrm{red}}},
 \qquad
 \mu_{\mathrm{red}}=\frac{m_em_h}{m_e+m_h}.
 \label{eq:dwc_optical_estimate}
\end{equation}
This estimate agrees well with the TDHF results. Increasing $d$ reduces both
$v_1(G)\propto e^{-Gd}$ and the electron density modulation, weakening
the interlayer restoring force. At fixed $d=8.2$~nm, the gap instead grows
by approximately $35\%$ between $R=5$ and $20$, consistent with the
sharpening density profiles and stronger interlayer locking.

The acoustic energies at M [Figs.~\ref{fig:tdhf_dwc_map}(b) and \ref{fig:tdhf_dwc_map}(c)] depend weakly on $d$
and decrease approximately as $R^{-1/2}$. For a common displacement of
the two layers, the mass density is $\rho_m=n(m_e+m_h)$, and isotropic
elasticity gives
\begin{equation}
 c_T^2=\frac{\mathcal G}{\rho_m},
 \qquad
 c_L^2=\frac{\mathcal B+\mathcal G}{\rho_m},
 \label{eq:dwc_elastic_velocities}
\end{equation}
with two-dimensional bulk and shear moduli $\mathcal B$ and
$\mathcal G$. Defining the finite-wave-vector estimates
$\widetilde c_\lambda=\omega_\lambda(q_{\min})/q_{\min}$, where
$q_{\min}=|\Gamma K|/6$, we find that
$\widetilde c_{L,T}\sqrt{1+R}$ vary by less than $10\%$ at $d=8.2$~nm
over $R=5$--$20$. The mass dependence is therefore dominated by inertia,
with comparatively weak changes in the effective stiffnesses.

The different responses to $d$ reflect the distinction between
compression and shear. A smooth neutral density fluctuation,
$\delta n_e=\delta n_h=\delta n$, has the Hartree energy density
\begin{equation}
 \delta\mathcal E_{\mathrm H}
 =[v_0(q)-v_1(q)]|\delta n|^2
 \underset{q\to0}{\longrightarrow}
 \frac{e^2d}{2\varepsilon_0\varepsilon_r}|\delta n|^2.
 \label{eq:dwc_capacitive_energy}
\end{equation}
Its contribution to the longitudinal velocity squared is
\begin{equation}
 c_{L,\mathrm H}^{\,2}
 =\frac{2n\lim_{q\to0}[v_0(q)-v_1(q)]}{m_e+m_h}
 =\frac{ne^2d}{\varepsilon_0\varepsilon_r(m_e+m_h)}.
 \label{eq:dwc_capacitive_velocity}
\end{equation}
At $R=10$ and $d=8.2$~nm, this gives $5.35\times10^4$~m/s, compared with
$\widetilde c_L\simeq5.05\times10^4$~m/s from TDHF. The observed increase
of $\widetilde c_L$ from $d=5.9$ to $10.9$~nm is approximately $28\%$,
close to the $36\%$ increase expected from $c_{L,\mathrm H}\propto\sqrt d$. This estimate
captures the capacitive scale; the full stiffness also contains kinetic,
exchange, and microscopic Hartree contributions, together with the
relaxation of the density profiles.

A transverse displacement produces no smooth density modulation to
linear order and therefore does not directly acquire this capacitive
contribution. The weak decrease of $\widetilde c_T$ with $d$ is consistent
with the reduced electron density modulation. Its stiffness nevertheless
includes the interlayer Hartree response of the microscopic crystal.
At finite momentum, optical softening further modifies the acoustic
dispersion through hybridization. For $R=10$, the optical feature near
one third of $\Gamma K$ decreases from approximately $2.6$~meV at
$d=6$~nm to $0.7$~meV at $d=11$~nm. Its avoided crossing with the TA
branch produces the shoulder in Fig.~\ref{fig:tdhf_cuts}(a). Close to the HC+EM boundary the hybridized TA mode near $\Gamma\mathrm{K}/2$ softens to zero: at $R=10$ its energy at $q=\Gamma\mathrm{K}/2$ falls from $0.44$~meV at $d=10.0$~nm to $0.14$~meV at $11.4$~nm, and it is imaginary from $11.8$~nm up to the upper edge of the dWC region at $12.7$~nm (dots in Fig.~\ref{fig:tdhf_dwc_map}). The dWC thus loses dynamical stability shortly before the HC+EM boundary, while the electron gap is still open.

\end{document}